\documentclass[%
reprint,
superscriptaddress,
amsmath,amssymb,amsthm,mathrsfs,
aps,
prl,
]{revtex4-1}

\usepackage{graphicx}
\usepackage{dcolumn}
\usepackage{bm}
\usepackage[usenames]{color}
\usepackage[ 
margin=0.75in,
]{geometry}
\usepackage[normalem]{ulem}

\graphicspath{{./figs/}}
\def\SNFA{Sr$_{1-x}$Na$_{x}$Fe$_2$As$_2$}

\def\Tc{$T_c$}
\def\Tr{$T_r$}
\def\Ts{$T_s$}
\def\CF{$C_4$}
\def\CT{$C_2$}

\def\rmid{$r_{\mathrm{mid}}$}
\def\dlt{$\delta$}

\begin{document}

\title{
Widespread orthorhombic fluctuations in the (Sr,Na)Fe$_2$As$_2$ family of superconductors
}

\author{Benjamin A. Frandsen}
\email{benfrandsen@byu.edu}
\affiliation{%
	Materials Sciences Division, Lawrence Berkeley National Laboratory, Berkeley, California 94720, USA.
}%
	\affiliation{ %
	Department of Physics and Astronomy, Brigham Young University, Provo, Utah 84602, USA.
} %

\author{Keith M. Taddei}
\affiliation{%
	Neutron Scattering Division, Oak Ridge National Laboratory, Oak Ridge, Tennessee 37831, USA.
}%

\author{Daniel E. Bugaris}
\affiliation{%
	Materials Science Division, Argonne National Laboratory, Argonne, Illinois 60115, USA.
}%

\author{Ryan Stadel}
\affiliation{ %
	Department of Physics, Northern Illinois University, DeKalb, Illinois 60115, USA.
} %
\affiliation{%
	Materials Science Division, Argonne National Laboratory, Argonne, Illinois 60115, USA.
}%

\author{Ming Yi}
\affiliation{ %
	Department of Physics, University of California, Berkeley, California 94720, USA.
} %
\affiliation{ %
	Department of Physics and Astronomy, Rice University, Houston, TX 77005, USA
} %

\author{Arani Acharya}
\affiliation{ %
	Department of Physics, University of California, Berkeley, California 94720, USA.
} %

\author{Raymond Osborn}
\affiliation{%
	Materials Science Division, Argonne National Laboratory, Argonne, Illinois 60115, USA.
}%

\author{Stephan Rosenkranz}
\affiliation{%
	Materials Science Division, Argonne National Laboratory, Argonne, Illinois 60115, USA.
}%

\author{Omar Chmaissem}
\affiliation{ %
	Department of Physics, Northern Illinois University, DeKalb, Illinois 60115, USA.
} %
\affiliation{%
	Materials Sciences Division, Argonne National Laboratory, Argonne, Illinois 60115, USA.
}%

\author{Robert J. Birgeneau}
\affiliation{ %
	Department of Physics, University of California, Berkeley, California 94720, USA.
} %
\affiliation{%
	Materials Science Division, Lawrence Berkeley National Laboratory, Berkeley, California 94720, USA.
}%
\affiliation{ %
	Department of Materials Science and Engineering, University of California, Berkeley, California 94720, USA.
} %

\begin{abstract}
We report comprehensive pair distribution function measurements of the hole-doped iron-based superconductor system \SNFA. Structural refinements performed as a function of temperature and length scale reveal orthorhombic distortions of the instantaneous local structure across a large region of the phase diagram possessing average tetragonal symmetry, indicative of fluctuating nematicity. These nematic fluctuations are present up to high doping levels ($x \gtrsim 0.48$, near optimal superconductivity) and high temperatures (above room temperature for $x = 0$, decreasing to 150~K for $x = 0.48$), with a typical length scale of 1--3~nm. This work highlights the ubiquity of nematic fluctuations in a representative iron-based superconductor and provides important details about the evolution of these fluctuations across the phase diagram.
\end{abstract}

\maketitle

A defining characteristic of the layered iron-based superconductors (FeSCs) is the presence of an electronic nematic phase in proximity to superconductivity~\cite{ferna;np14,hoson;pc15,si;nrm16}. In analogy to nematic phases in liquid crystals, nematic order in FeSCs lowers the \CF\ rotational symmetry present in the system at high temperature to \CT\ at low temperature. This symmetry breaking manifests itself through anisotropic electronic properties~\cite{chuan;s10,chu;s10,chu;s12}, a tetragonal-to-orthorhombic structural phase transition~\cite{johre;zk11}, the formation of stripe-type magnetic order~\cite{dai;np12}, and a splitting of $d$ orbitals in the electronic band structure~\cite{yi;qm17}. The nematic order present in the parent compounds of most FeSCs can be suppressed by chemical substitution and/or pressure, with the highest superconducting \Tc\ typically appearing near the point of complete suppression of the nematic phase. Consequently, nematicity is believed to be intimately related to the still elusive superconducting mechanism in FeSCs~\cite{ferna;np14,shiba;arocmp14,chen;nsr14,leder;prl15,bohme;crp16,kuo;s16,matsu;nc17,ferna;rpp17}.

Research efforts into nematicity have recently expanded their focus to include not only the nematic phase itself, but also nematic fluctuations present outside the region of static nematicity. The hope is that such fluctuations may reveal important details about the origin of nematicity and ultimately superconductivity. Signatures of high-temperature nematicity above the static nematic phase have in fact been observed in numerous iron pnictide and chalcogenide systems with a variety of experimental probes~\cite{rosen;np14,iye;jpsj15,galla;prl16,hosoi;pnas16,massa;pnas16,kretz;np16,liu;prl16,xu;prb16,luo;qm17,palms;prb17,wang;prb17,baek;nc18}, indicating that an effort to characterize these nematic fluctuations systematically within and between phase diagrams of representative FeSCs holds great promise.

According to current understanding~\cite{chu;s12}, nematic order drives the orthorhombic phase transition; hence, nematic fluctuations will likewise cause fluctuating orthorhombic distortions of the lattice. For simplicity, we shall use the terms ``nematic fluctuations'' and ``orthorhombic fluctuations'' interchangeably. The pair distribution function (PDF) method of analyzing x-ray and neutron scattering data~\cite{egami;b;utbp12} has recently been shown to be an effective probe of nematic fluctuations in the \SNFA\ system, a representative hole-doped FeSC family~\cite{frand;prl17}. This technique involves Fourier transforming the total scattering data into real space to obtain the pairwise atomic correlations. Because both Bragg and diffuse scattering are included in the Fourier transform, the PDF pattern is sensitive to the local atomic structure, even if it differs from the average crystallographic structure. As such, the PDF technique is an ideal probe of short-range structural distortions such as those associated with nematic fluctuations. The PDF analysis in Ref.~\onlinecite{frand;prl17} revealed short-range orthorhombic distortions on a length scale of $\sim$2~nm well into the high-temperature tetragonal phase for two underdoped samples and in the magnetic \CF\ phase that appears in this and similar hole-doped iron pnictides~\cite{allre;prb15,allre;np16,tadde;prb16,chris;prl18}. The ability to investigate the length scale of these nematic fluctuations makes the PDF approach particularly valuable.

\begin{figure}
	\includegraphics[width=70mm]{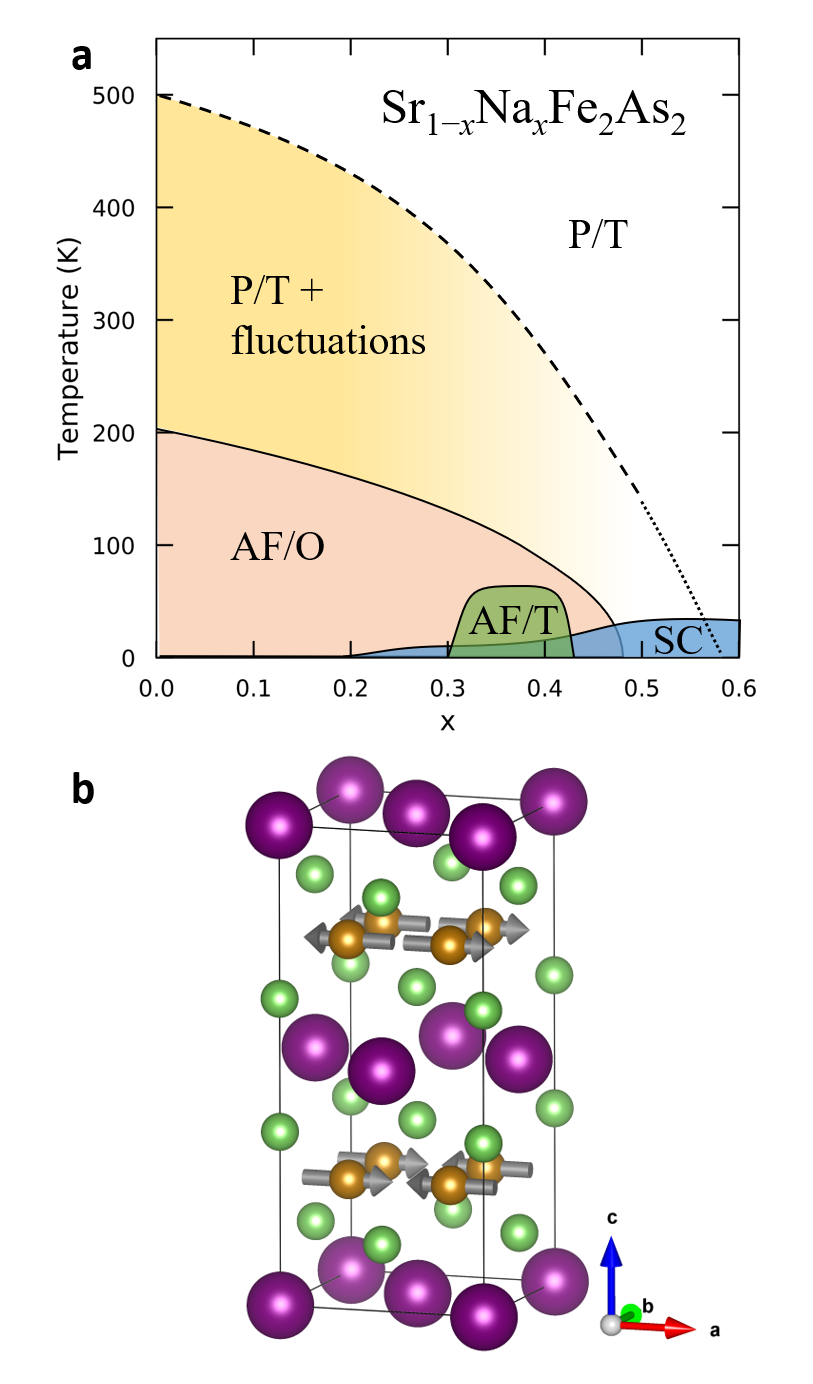}
	\caption{\label{fig:PD}   Phase diagram and structure of \SNFA. (a) Temperature versus composition phase diagram of \SNFA, showing the paramagnetic tetragonal (P/T), antiferromagnetic orthorhombic (AF/O), antiferromagnetic tetragonal (AF/T), and superconducting (SC) phases, along with a large region of nematic fluctuations shaded in yellow. Solid lines denote true phase transitions, the dashed line above the yellow region represents an estimated onset temperature of nematic fluctuations, and the dotted line that closes the yellow region is an attempt at a reasonable extrapolation beyond the region of the phase diagram for which we have data. (b) Crystal structure of \SNFA\ in the orthorhombic setting. Gold, green, and purple spheres represent iron, arsenic, and strontium/sodium atoms, respectively. The arrows show the stripe-type magnetic order.}
	
\end{figure}

Here, we extend the previous PDF work to a much larger region of the \SNFA\ phase diagram. We studied several compositions ranging from the parent compound with $x = 0$ to a nearly optimally doped compound with $x = 0.48$, for which the long-range orthorhombic structural transition is completely absent. The main result is our observation of local orthorhombic distortions with a typical length scale of 1-3~nm in all measured compounds up to remarkably high temperatures, with a maximum of approximately 500~K for $x = 0$ and a gradual decline to $\sim$150~K for $x = 0.48$. The magnitude and length scale of the local orthorhombicity likewise decrease monotonically with $x$. Taken together, these measurements map out in great detail a large region of fluctuating, short-range nematicity in the \SNFA\ phase diagram [see Fig.~\ref{fig:PD}(a)], with significant nematic fluctuations remaining even near the optimally doped region.
These results further establish the widespread presence of significant nematic fluctuations in FeSCs and provide important new information about their length and temperature scales in \SNFA. 

We studied powder specimens of \SNFA\ with $x = 0$, 0.12, 0.27, 0.29, 0.34, 0.45, and 0.48. As seen in the phase diagram, the samples with $x \le 0.29$ undergo a tetragonal-to-orthorhombic structural phase transition at \Ts\ and remain orthorhombic for all lower temperatures, whereas the $x = 0.34$ sample exhibits reentrant tetragonal symmetry below \Tr~$<$~\Ts. The sample with $x = 0.45$ lies beyond the \CF\ dome and undergoes a single orthorhombic transition at \Ts, and the $x = 0.48$ sample retains average tetragonal symmetry at all temperatures and exhibits no magnetic order. All structural and magnetic transitions are first order. A superconducting ground state exists for compositions with $x \gtrsim 0.2$. Detailed characterization of these samples can be found in Ref.~\onlinecite{tadde;prb16}. For reference, Fig.~\ref{fig:PD}(b) displays the crystal structure in the orthorhombic setting and the corresponding stripe-type magnetic order. The orthorhombic distortion causes $a$ to become larger than $b$.

The samples with $x = 0$, 0.29, 0.34, 0.45, and 0.48 were measured on the NOMAD beamline~\cite{neufe;nimb12} at the Spallation Neutron Source (SNS) of Oak Ridge National Laboratory (ORNL). PDF data were generated from the total scattering data with $Q_{\mathrm{max}}$ = 36~\AA$^{-1}$ using the ADDIE software suite~\cite{mcdon;aca17}. Additional samples with $x = 0$, 0.12, and 0.27 were measured on the XPD beamline of the National Synchrotron Light Source II (NSLS-II) at Brookhaven National Laboratory (BNL). The data were reduced using the xPDFsuite software~\cite{yang;arxiv15} with $Q_{\mathrm{max}}$ = 25~\AA$^{-1}$. On both beamlines, data were collected on dense temperature grids between $\sim$5~K and 300~K, with typical temperature steps of 10-15~K on NOMAD and 6~K on XPD. Structural refinements were performed using the PDFgui program~\cite{farro;jpcm07} and the Diffpy-CMI suite~\cite{juhas;aca15}. We note that neither beamline employs energy analysis of the scattered particles. As such, the PDF patterns contain meaningful information from inelastically scattered particles within some effective energy window, which is estimated to be tens of meV on NOMAD and effectively infinite on XPD. The neutron data collected on NOMAD therefore probe structural correlations on time scales of approximately 10$^{-13}$~s or longer, and the x-ray data represent the true instantaneous structure. In the present work, there seems to be little discernible difference between the two.

We first present our analysis of the neutron PDF data. To determine the atomic structure as a function of both length scale and temperature, we performed an extensive series of fits for each composition measured. For a given temperature, we refined the \textit{Fmmm} orthorhombic structural model against a sliding 20-\AA\ data window ranging from [1.5~\AA\ - 21.5~\AA] to [30.5~\AA\ - 50.5~\AA] in 1-\AA\ steps, resulting in 30 fits per temperature. The results of each refinement represent the best-fit atomic structure on the length scale set by the data window. This was then repeated for each temperature at which PDF data were collected. Additional details about the refinements are included in the Supplementary Information~\footnote{See Supplemental Material at ... for further information about the refinements and figures showing representative fits.}.

\begin{figure}
	\includegraphics[width=60mm]{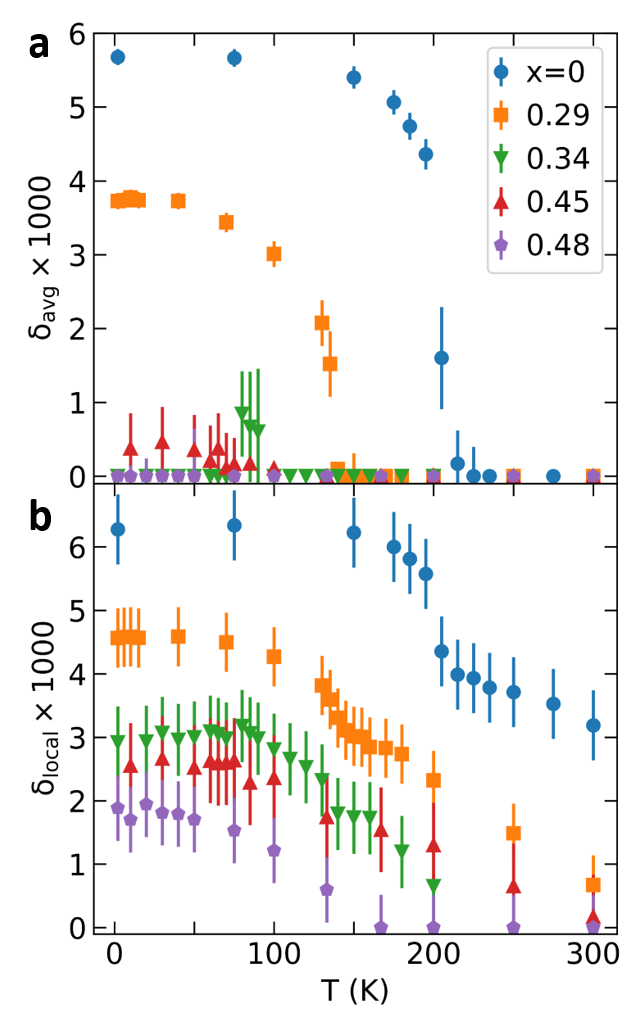}
	\caption{\label{fig:dltResults}   Refined orthorhombicity \dlt\ extracted from fits to neutron PDF data. (a) Results for a long fitting range of 30.5 - 50.5 \AA, showing the expected behavior for the average structure determined from Rietveld refinements. (b) Results compiled from shorter-range fits (1.5 - 21.5~\AA, 2.5 - 22.5~\AA, 3.5 - 23.5~\AA) showing enhanced orthorhombicity up to significantly higher temperatures than observed in the long-range structure.}
	
\end{figure}

The output of this fitting procedure is a comprehensive set of structural parameters as a function of length scale, temperature, and chemical composition, providing a rich and detailed view of the local structure across the \SNFA\ phase diagram. The structural parameter most relevant to nematic fluctuations is the orthorhombicity, $\delta = (a-b)/(a+b)$. In Fig.~\ref{fig:dltResults}(a), we display the refined orthorhombicity as a function of temperature for fits conducted over a long fitting range of 30.5 - 50.5~\AA. 
All samples show an abrupt development of nonzero orthorhombicity at the expected structural phase transition temperature, except for the sample with $x = 0.48$ which lies beyond the \CT\ region and does not transition to an average orthorhombic structure at any temperature. The refined values of the orthorhombicity agree quantitatively with the average crystallographic structure determined from traditional Rietveld analysis in Ref.~\onlinecite{tadde;prb16}, indicating that the PDF data on a length scale of 30.5 - 50.5~\AA\ show no detectable difference from the average structure.

The refinements conducted over shorter fitting ranges yield strikingly different results. Fig.~\ref{fig:dltResults}(b) displays the orthorhombicity averaged over the three shortest fitting ranges. Compared to the long-range structure, the orthorhombicity of each sample is enhanced at low temperature and persists to much higher temperatures far in excess of the transition temperature. Particularly notable is the $x = 0.48$ sample, which shows significant orthorhombicity up to $\sim$150~K, even though the average crystallographic structure remains tetragonal at all temperatures. These results demonstrate that the instantaneous local structure of \SNFA\ remains orthorhombically distorted well into the high-temperature paramagnetic tetragonal phase, even for superconducting samples wholly outside the doping region undergoing a long-range \CT\ transition.

As discussed in Ref.~\onlinecite{frand;prl17}, we expect these short-range orthorhombic distortions to be fluctuating dynamically on a time scale between 10$^{-13}$ and 10$^{-7}$~s; hence, we refer to them as nematic fluctuations. The upper bound of this time window is based on the absence of any observed symmetry breaking outside the \CT\ phase in nuclear magnetic resonance measurements of \SNFA~\cite{allre;np16}, while the lower bound comes from the estimated effective energy window of 10~meV for the neutron PDF measurements. For the samples with $x$~=~0, 0.12, and 0.27 measured with x rays, the time scale is truly instantaneous. The dynamic and short-ranged nature of these nematic fluctuations renders them undetectable by techniques lacking sensitivity to nanometer-scale structure on a time scale of 10$^{-13}$ to 10$^{-7}$~s.

\begin{figure*}
	\includegraphics[width=160mm]{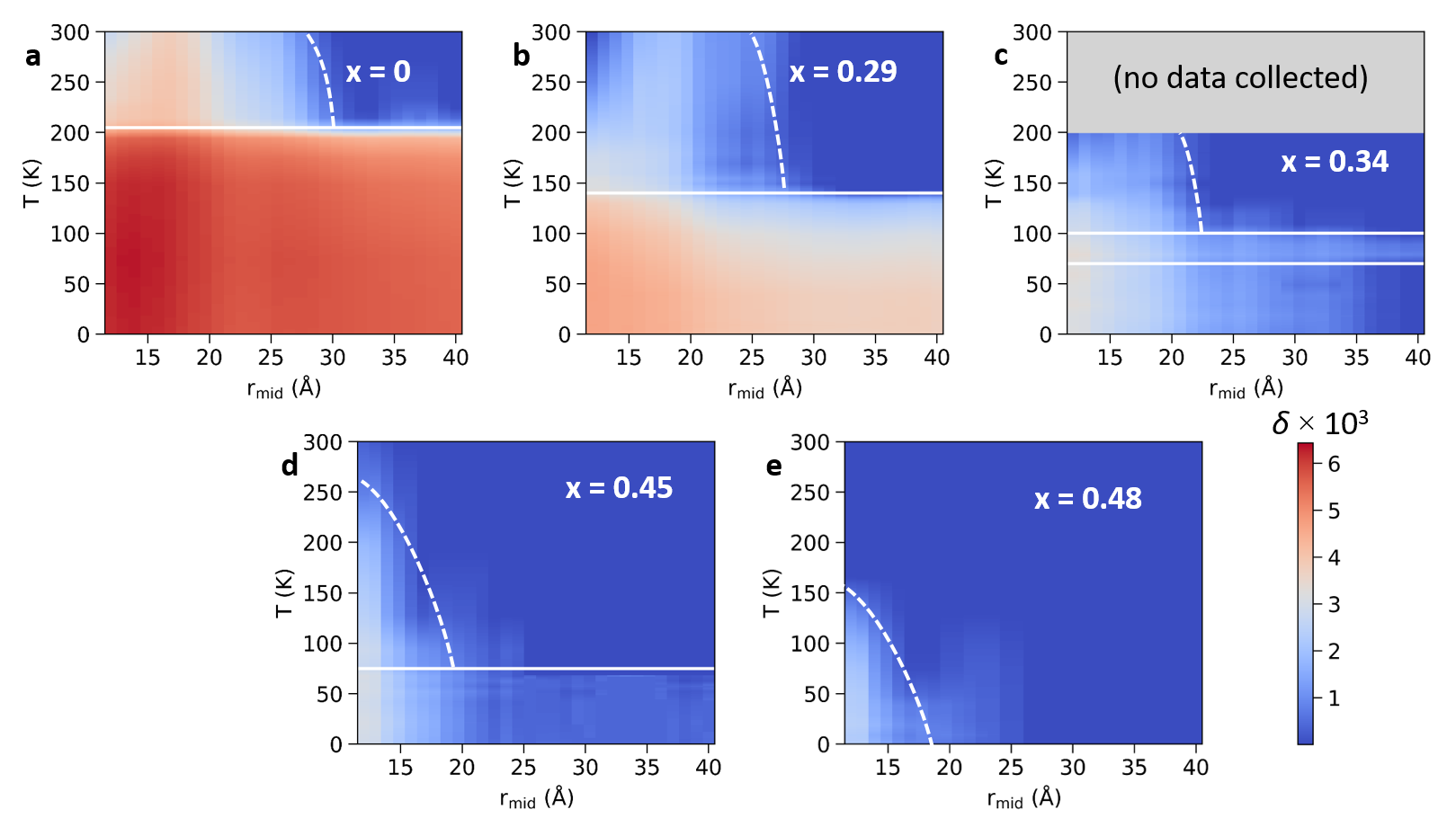}
	\caption{\label{fig:dltMaps}   False color plots of the refined orthorhombicity determined from neutron PDF analysis of samples of \SNFA\ with $x = 0$ (a), 0.29 (b), 0.34 (c), 0.45 (d), and 0.48 (e). The horizontal axis represents the midpoint of the fitting window (see main text), the vertical axis temperature. White horizontal lines indicate structural phase transitions that occur in the average structure for each compound. The dashed white curves are guides to the eye that loosely define regions of nonzero local orthorhombicity inside the paramagnetic tetragonal phase, a manifestation of nematic fluctuations. Linear interpolation has been used between neighboring $T$ and \rmid\ points.}
	
\end{figure*}

To determine how the local structure evolves into the average stucture, we present in Fig.~\ref{fig:dltMaps} false color plots of \dlt\ from our comprehensive $T$- and $r$-dependent refinements.
Temperature is displayed on the vertical axes, the midpoint of each fitting range on the horizontal axes. The brightness of the plots scales with the orthorhombicity, as indicated by the color bar. All five plots use a shared color bar to provide a better indication of the overall evolution of the magnitude of the orthorhombic distortion. The horizontal white lines demarcate the tetragonal and orthorhombic phases observed in the average structure.

Focusing initially on the plot for $x = 0$, we observe a bright block of intensity for all values of \rmid\ below the white line marking \Ts~=~205~K, consistent with a long-range orthorhombic distortion of the crystal structure. Above \Ts, however, nonzero orthorhombicity exists only on the low-$r$ side of the plot, gradually decreasing to zero beyond $r_{\mathrm{mid}}\approx 30$~\AA\ (the dashed white curves are guides to the eye). This is a clear demonstration of the short-range nature of the local orthorhombic distortion at high temperature, indicating a length scale of approximately 3~nm. Fig.~\ref{fig:dltMaps}(b) shows qualitatively similar behavior for the $x = 0.29$ sample, albeit with overall lower values of \dlt\ and a somewhat shorter length scale for the local distortion at high $T$.

The results for $x = 0.34$, shown in Fig.~\ref{fig:dltMaps}(c), are notable due to the reentrant \CF\ phase that sets in below 70~K. As a result, the average structure is orthorhombic only between $\sim$100 and 70~K, seen as the narrow strip of intensity between the two horizontal white lines. On the high $r$ side of the plot, no intensity exists outside this temperature region, in accordance with expectations for the tetragonal average structure. Contrastingly, the low $r$ region of the plot shows significant intensity at all temperatures. This demonstrates that the local structure remains orthorhombically distorted in the entirety of the magnetic \CF\ phase and well into the paramagnetic \CF\ phase at higher temperature, fully consistent with the results reported previously~\cite{frand;prl17}.

Finally, we consider the color maps for $x$~=~0.45 and 0.48, shown in Fig.~\ref{fig:dltMaps}(d) and (e), respectively. Both show significant local orthorhombicity on a $\sim$15~\AA\ length scale in the nominally tetragonal phase, reaching up to $\sim$250~K for $x = 0.45$ and 150~K for $x = 0.48$. The orthorhombic length scale in both compounds increases slightly as the temperature is lowered. For $x = 0.45$, the orthorhombicity below $\sim$75~K remains nonzero but small for long $r$. No significant orthorhombicity exists at long $r$ for $x = 0.48$, as expected since it is known to remain in the tetragonal phase at all temperatures.

The x-ray PDF data collected for the samples with $x = 0$, 0.12, and 0.27 were analyzed similarly. The results for $x = 0$ are in good agreement with the neutron data collected on the same sample, and the results for $x = 0.12$ and 0.27 interpolate well between $x = 0$ and the higher dopings. We found that the refined values of the orthorhombicity are systematically slightly lower for the x-ray data than for the neutron data, which may be due to differences in the real-space resolution between the two types of experiment. However, the $r$- and $T$-dependent trends in the x-ray data are robust and fully consistent with those in the neutron data.

\begin{figure}
	\includegraphics[width=60mm]{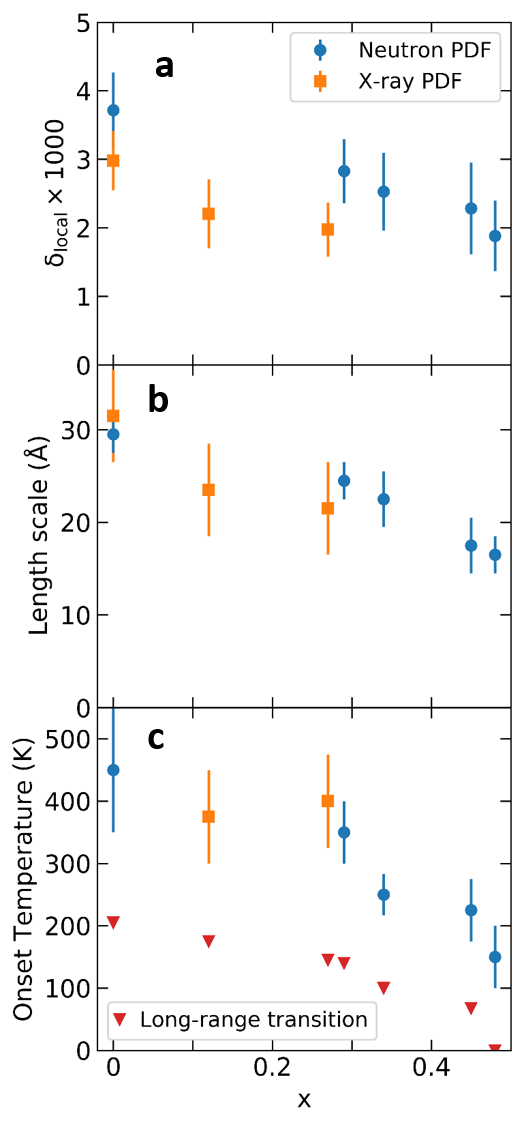}
	\caption{\label{fig:trends}   Composition-dependent trends in the (a) magnitude, (b) length scale, and (c) onset temperature at which the local orthorhombic distortion in the paramagnetic tetragonal phase becomes observable in the PDF analysis. A monotonic decrease in all three quantities is observed with increasing doping $x$. Blue circles come from analysis of the neutron PDF data, orange squares from the x-ray data. The results in (a) and (b) were obtained from data sets measured at approximately 1.2~$\times$~\Ts. The red triangles in (c) mark the long-range transition temperature from the paramagnetic tetragonal to the antiferromagnetic orthorhombic phase.}
	
\end{figure}

Our PDF analysis establishes the presence of significant nematic fluctuations in the high-temperature tetragonal phase of \SNFA\ up to doping levels of at least $x = 0.48$, close to the region of optimal superconductivity. We can extract overall trends in these nematic fluctuations across the phase diagram by considering the doping dependence of the PDF results. In Fig.~\ref{fig:trends}(a), we show the magnitude of the local orthorhombicity in the paramagnetic tetragonal phase for each composition measured. 
The values of \dlt\ shown were extracted from the short-range fits to the PDF patterns collected at temperatures of approximately 1.2~$\times$~\Ts\ (for $x = 0.48$, we chose the cryostat base temperature of 2~K). Blue circles originate from neutron data, orange squares from x-ray data. Ignoring the previously mentioned offset of the x-ray results slightly below the neutron results, the orthorhombicity clearly tends to decrease with increasing doping level $x$, consistent with the reduction of the orthorhombic distortion in the average structure as $x$ increases. In Fig.~\ref{fig:trends}(b), we display the typical orthorhombic length scale at 1.2~$\times$~\Ts\ for each composition, determined as the midpoint of the shortest fitting range for which the orthorhombicity refines to zero within the parameter uncertainty. A similar decreasing trend is observed. Finally, Fig.~\ref{fig:trends}(c) illustrates the approximate onset temperature of the short-range orthorhombic distortion (i.e., the temperature at which it becomes observable in the PDF analysis) as a function of $x$. The onset temperature is remarkably high for all compositions measured, typically greater than twice the long-range \CT\ transition temperature. For $x = 0.45$ and 0.48, the onset temperature was determined from the data directly, since measurements were conducted at sufficiently high temperatures for \dlt\ to refine to zero. However, the other samples remained locally orthorhombic at the highest measured temperatures, so we estimated the onset temperature by linearly extrapolating the refined values of \dlt\ to zero. Although this analysis is not particularly precise [as reflected by the large error bars in Fig.~\ref{fig:trends}(c)], it suffices for an approximate measure of the relevant temperature scale for the local orthorhombic structure.  

In summary, the PDF analysis reported here reveals the presence of high-temperature nematic fluctuations in \SNFA\ and clarifies their essential characteristics. These fluctuations exist on a length scale of 1--3~nm, persist up to remarkably high temperatures in excess of 2\Ts, and extend to doping levels well into the superconducting dome and entirely outside the parameter space with a long-range orthorhombic structural transition. The large area of temperature-composition parameter space supporting this fluctuating nematicity is represented by the yellow shaded region in Fig.~\ref{fig:PD}. We do not expect these results to be strongly affected by the disorder introduced by Na substitution, since it occurs in the Sr plane, mitigating the effect on the Fe plane, and because the structural-magnetic transitions are first order~\footnote{Both hole- (e.g. \SNFA)  and electron- (e.g. Ba(Fe$_{1-x}$Co$_x$)$_2$As$_2$) doped Fe-based superconducting materials typically will have some level of dopant disorder, which will increase in importance with increasing dopant concentration. These dopant compositional fluctuations will act like a random field on the orthorhombic structural order parameter. For the hole-doped systems, such as that studied here, the primary effect will be to round the otherwise sharp first order structural-magnetic phase transition. For typical electron-doped systems with separated structural and magnetic transitions, the random fields will have more drastic effects. Specifically, the random fields will prevent a true divergence of the orthorhombic structural correlation length due to metastability effects and, concomitantly, cut off the divergence of the stripe magnetic correlation length at the structural value [see Fisher, \textit{Phys. Rev. Lett.} \textbf{56}, 416 (1986) and Birgeneau \textit{et al.}, \textit{Phys. Rev. Lett.} \textbf{75}, 1198 (1995)].  Thus, there can be no true second-order structural and magnetic transitions in systems such as Ba(Fe,Co)$_2$As$_2$ due to the dominant role of metastability in random field Ising model (RFIM) systems. These metastability effects could be most important near any putative quantum critical point and, specifically, could prevent accessing equilibrium quantum critical behavior due to the RFIM metastability.}. The observation of nematic fluctuations up to such high temperatures, while perhaps initially surprising, is consistent with other experimental results on iron pnictide systems, such as elastic shear modulus data showing anomalies still present at room temperature~\cite{bohme;crp16}. Our findings are also reminiscent of a recent report of local orthorhombicity at high temperature and doping in Na(Fe,Ni)As inferred from neutron diffraction data~\cite{wang;nc18}, although the techniques used in that work probe structural correlations on a longer time scale than applies to the current PDF data.

The high-temperature nematic fluctuations revealed here may have a primarily two-dimensional character~\cite{wilso;prb10}, which could help explain the high onset temperature. Diffuse scattering studies of single crystals would be valuable to investigate this further. Measurements of the stripe-type magnetic correlation length in the paramagnetic phase would also provide an important comparison with the local orthorhombic length scale reported here. The observation of local orthorhombic distortions persisting to doping levels beyond the \CT\ dome reported here is a rare example of direct experimental evidence that nematic degrees of freedom are still active near optimal superconductivity. This is an important factor for various theoretical proposals suggesting that such fluctuations can induce or enhance superconductivity, such as in Refs.~\onlinecite{leder;prl15,chubu;prx16,leder;pnas17}. Overall, this work highlights the ubiquity of nematic fluctuations in a canonical hole-doped FeSC system and invites similar studies in other representative families. A thorough characterization of nematic fluctuations across multiple families of FeSCs promises to yield important insights into the origin of the nematic phase and its role in iron-based superconductivity.


\textbf{Acknowledgements}

We thank Steve Kivelson, Ian Fisher, Amnon Aharony, and Dung-Hai Lee for valuable conversations about these results. Work at Lawrence Berkeley National Laboratory was funded by the U.S. Department of Energy, Office of Science, Office of Basic Energy Sciences, Materials Sciences and Engineering Division under Contract No. DE-AC02-05-CH11231 within the Quantum Materials Program (KC2202). Work at the Materials Science Division at Argonne National Laboratory was supported by the US DOE, Office of Science, Materials Sciences and Engineering Division. Use of the SNS, ORNL, was sponsored by the Scientific User Facilities Division, BES, US DOE. Use of the National Synchrotron Light Source II at Brookhaven National Laboratory, was supported by DOE-BES under Contract No. DE-SC0012704.

\end{document}